\PassOptionsToPackage{hyphens}{url}
\documentclass[conference]{IEEEconf}
\IEEEoverridecommandlockouts
\usepackage{cite}
\usepackage{amsmath,amssymb,amsfonts}
\usepackage{algorithmic}
\usepackage{graphicx}
\usepackage{svg}
\svgpath{{Figures/}}
\usepackage{textcomp}
\usepackage{xcolor}
\usepackage{hyperref}
\hypersetup{hidelinks}
\usepackage{pdfpages}
\usepackage{url}
\usepackage[sort&compress, nameinlink]{cleveref}

\crefname{equation}{}{} 
\Crefname{equation}{}{} 

\usepackage{subfig}
\usepackage{dsfont}
\usepackage{units}
\usepackage[nolist]{acronym}

\newtheorem{thm}{Theorem}
\newtheorem{prob}[thm]{Problem}
\def\BibTeX{{\rm B\kern-.05em{\sc i\kern-.025em b}\kern-.08em
    T\kern-.1667em\lower.7ex\hbox{E}\kern-.125emX}}

\newif\ifmargincomments 
\margincommentstrue
\usepackage{marginnote}

\newif\ifextendedversion 
 \extendedversiontrue

\ifmargincomments

\else

\fi 

\begin{document}

\title{ \bf{Multi-objective Electric Vehicle Charging Strategies at Energy Hubs with Justice-informed Profit Distribution among Users}}
\author{Twan Stolwijk$^\star$, Diego Fernández-Zapico$^\star$, Alessandro Piazza, Mauro Salazar
	\thanks{Control Systems Technology section, Department of Mechanical Engineering, Eindhoven University of Technology, Eindhoven, The Netherlands. E-mails:  {\tt\small \{t.m.stolwijk, d.fernandez.zapico, a.p.piazza1, m.r.u.salazar\}@tue.nl}.\newline
	$^\star$The first two authors contributed equally to this paper.}%
}

\begin{acronym}
	\acro{EVCS}{electric vehicle charging station}
	\acro{EV}{electric vehicle}
	\acro{V2G}{vehicle-to-grid}
	\acro{MPC}{model predictive control}
	\acro{GHG}{green house gas}
	\acro{BESS}{battery energy storage system}
	\acro{CAPEX}{capital expenditure}
	\acro{OPEX}{operational expenditure}
	\acro{BESS}{battery energy storage system}
\end{acronym}

\maketitle
\thispagestyle{plain}
\pagestyle{plain}


\begin{abstract}
This paper studies potential energy cost savings in electric vehicle (EV) charging hubs when users allow cooperative vehicle-to-grid (V2G), and how such savings should be redistributed.
To this end, we present multi-objective operational strategies and justice-informed profit sharing methods for a charging energy hub equipped with a battery storage system and solar panels.
Specifically, we first frame the dynamic energy hub operational problem aiming to minimize a combination of grid costs and emissions, accounting for time-varying prices and energy mix as well as storage system and EV battery degradation, which is especially relevant when considering V2G operations. We parse the problem as a second-order conic program that can be efficiently solved to global optimality.
Our real-world case-study shows the importance of flexibility for hub operation: Equipping a charging station with a stationary battery and leveraging dynamic V2G can significantly reduce the total costs and emissions compared to conventional settings without a battery or dynamic charging by more than 10\%.
To redistribute these cost benefits to users, we compare standard Nash bargaining with a justice-informed sufficientarian version which caps charging prices to safeguard adequate access to affordable charging.
Our results show that the sufficientarian approach better accounts for inequalities in flexibility capital without significantly affecting aggregate costs, highlighting how justice-informed pricing only marginally affects total revenues.
\end{abstract}


\section{Introduction}
\begin{figure}
	\centering
	\def\svgwidth{0.9\linewidth}
    \input{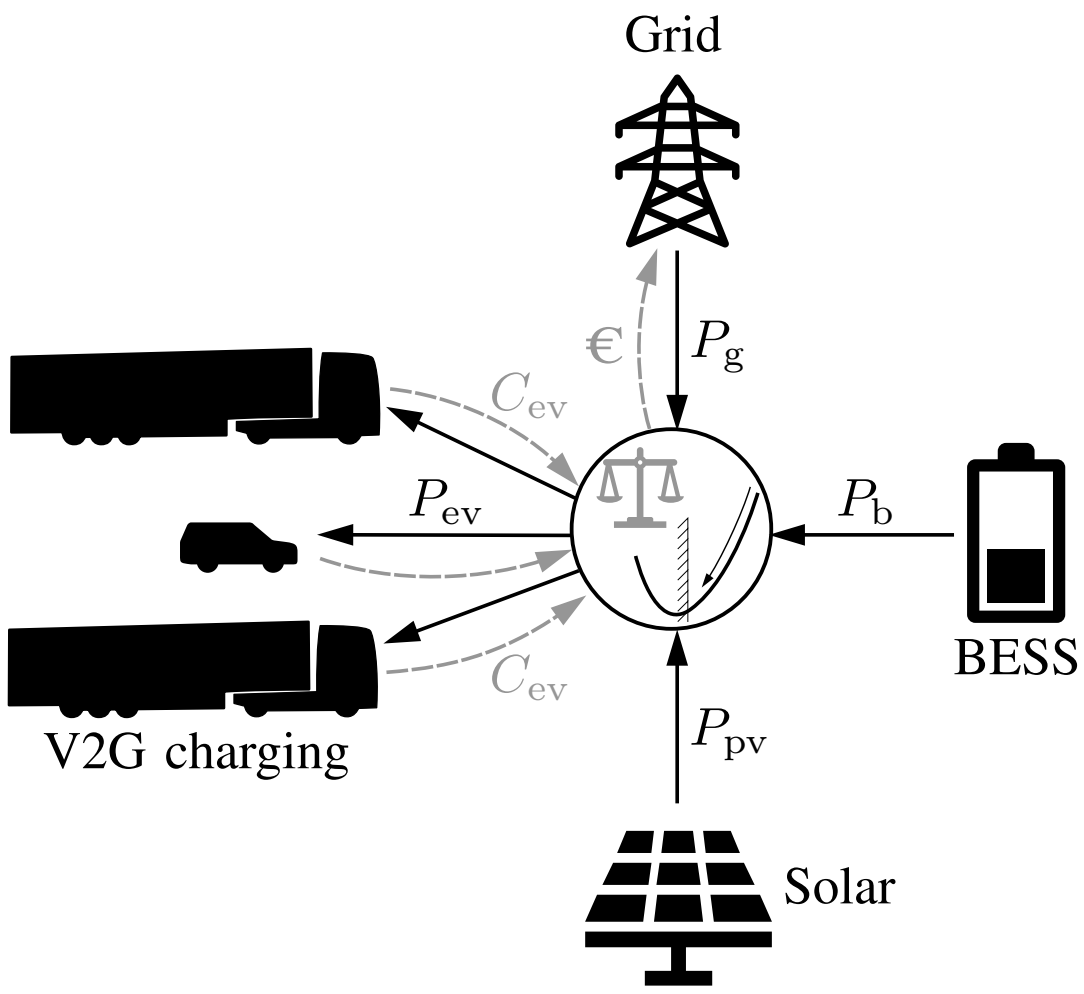}
	\caption{Schematic of the optimal power flow by the controller for the hub components and fair savings sharing from smart charging among users.}
	\label{fig:Interconnection_hub}
\end{figure}

Climate change necessitates a sustainable transition to renewable energy, with electrification playing a central role. The energy transition has rapidly increased in the number of \acp{EV} and the share of energy generated through intermittent renewable sources ~\cite{IEA2024b}. This has placed increased stress on an electricity grid that was not designed for the irregular electricity supply from an energy mix increasingly dominated by renewable energy sources and does not coincide with daily demand profiles~\cite{MliloBrownEtAl2021}. Charging stations supported by on-site renewable generation, a \ac{BESS} and vehicle-to-grid (V2G) charging as shown in Fig.~\ref{fig:Interconnection_hub} can help overcome these challenges by providing additional flexibility~\cite{AnwarMuratoriEtAl2022}. 

At the same time, the energy transition should not reinforce economic and social inequalities. 
Unlike conventional refuelling, EV charging sessions take significantly longer, making users’ ability to choose when and where to charge increasingly important. Moreover, as EV adoption grows, a larger share of users will lack access to home charging~\cite{GeSimeoneEtAl2021}, further limiting their charging options. This makes flexibility capital, understood as users’ capacity to adapt their energy use in light of the constraints and demands of everyday life, a key dimension of social justice in the energy transition~\cite{Libertson2022}. 
However, this capacity is not equally distributed. Disadvantaged households face greater constraints on when and where they can charge their vehicles, limiting their ability to take advantage of cheaper charging opportunities and exposing them to charging disadvantage~\cite{MalabananSauriLavieriEtAl2025}. 

This work combines cost- and emission-optimal control of an EV charging hub with equitable mechanisms for distributing the benefits of smart charging among users.

\subsubsection*{Related literature}
This paper pertains to two research streams: cost functions for controlling energy hubs and different cost distribution methods among participants.

In a recent literature review, authors compare optimal \ac{EVCS} power management strategies~\cite{GhanbariMotlaghOladigboluEtAl2025}. Optimisation-based controllers for \ac{EVCS} operation include \ac{MPC}~\cite{LeFlochBansalEtAl2019, PolancoVasquezLopezRedondoEtAl2023, CeustersRodriguezEtAl2021}, which has the advantage of providing optimal actions over a prediction horizon while enforcing constraint satisfaction. 
The objective functions of these optimisation-based methods may consider different goals. Most commonly, controllers minimise the total economic cost of charging station owners or users~\cite{FernandezZapicoHofmanEtAl2025, PatilKalkhambkar2020}. In other cases, cost functions account for component amortisation costs \cite{ZhangHoreshEtAl2023}, grid power stabilisation \cite{GanTopcuEtAl2013, AounAddaEtAl2024} or  \ac{GHG} emissions \cite{KarmakerHossainEtAl2023, HoehneChester2016}. 
Other work explores the multi-objective optimisations using combinations of the aforementioned cost functions \cite{PolancoVasquezLopezRedondoEtAl2023, LeeMadanat2017, AldosariAliEtAl2024}. However, as far as authors are aware, no previous work compares the Pareto fronts achievable by different (smart) EV charging strategies for multi-objective optimal control of a charging energy hub.

Many works use a utilitarian cost function to minimise the total operating costs, which does not consider the social implications for individual EV users. Some papers address this through the use of heuristics, which have no optimality guarantees and, in most cases, lack social substantiation \cite{AounAddaEtAl2024, FrolikHines2012}. 
Another approach is cost division based on game theory. In \cite{PaudelWolfEtAl2024}, authors use Nash bargaining to find a profit distribution between a fast charging hub and a nearby BESS.
Furthermore, in \cite{BahramiToulabiEtAl2018} Nash bargaining determines the optimal operation of an energy network with multiple energy generators.
Another paper \cite{XieChen2024} proposes a carbon emission allocation scheme based on the marginal contribution to emissions of energy storage in an energy market.
Other work uses a dynamic pricing scheme for a charging station \cite{LimmerDietrich2018}. They aim to minimise the perceived unfairness due to price differences between different users with similar energy requests. 
In \cite{ShilovElokdaEtAl2026}, authors present a framework for determining the applicability and comparability of different social cost functions. 

In conclusion, as far as the authors are aware, there is no previous work on the Pareto front of objective functions considering price and emissions for a charging energy hub operated under multiple smart charging strategies. Additionally, the use of a socially substantiated allocation method to fairly divide the savings generated by smart charging among cooperating users does not exist yet. 

\subsubsection*{Statement of contributions}
The main contributions in this paper are twofold: First, we present a comparison of multiple (smart) charging strategies for an energy hub. We show that through optimal control the operating costs and \ac{GHG} emissions for EV charging can be significantly reduced by leveraging dynamic charging and V2G.
Secondly, we propose two game-theoretic methods to distribute these savings among users and the operator. Beyond implementing Nash bargaining, we introduce a sufficientarian extension that caps the charging price to ensure that lower-income households with limited scheduling flexibility retain sufficient access to affordable charging, whilst not significantly affecting the overall revenues.

\subsubsection*{Organisation} The remainder of this paper is structured as follows:
Section~\ref{ch: energy hub model} presents a model for the EVCS and introduces the different EV charging strategies. This is subsequently formulated into a multi-objective optimisation problem that aims at minimising combinations of economic cost and GHG emissions in Section~\ref{ch:Optimal_hub_operation}. 
In Section~\ref{ch:Game theory} we first propose a cost allocation method based on Nash bargaining. Furthermore, we present a sufficientarian approach to reduce the inequality caused by users’ temporal inflexibilities. 
Section~\ref{ch:Case_study} presents a case study for a charging energy hub on a university campus. 
Finally, conclusions are drawn in Section~\ref{ch:conclusion} together with an outlook on future work.

\section{Charging Energy Hub Model}
\label{ch: energy hub model}
This section presents the model of the energy hub used for the multi-objective optimisation, accounting for component degradation and amortisation costs. We consider the EVCS from Fig.~\ref{fig:Interconnection_hub}, which includes power flows from the grid connection ($P_\mathrm{g}$), the \ac{BESS} ($P_\mathrm{b}$), solar panels ($P_\mathrm{pv}$) and the EV chargers ($P_\mathrm{ev}$).  A central controller determines optimal component power flow to meet the charging load as
\begin{equation}
    \label{eq:Hub_Pwrbalance}
    P_{\mathrm{ev}}(t) = P_{\mathrm{g}}(t) + P_{\mathrm{pv}}(t) + P_{\mathrm{b}}(t).
\end{equation}

\subsection{BESS}
We model the BESS using the approach described in \cite{ClementeSalazarEtAl2024, VerbruggenSalazarEtAl2019}, with the relaxed second order conic relaxation of the quadratic model for the battery losses as
\begin{equation}
    \label{eq:battery_loss_ib}
    P_{\mathrm{ib}}(t) - P_{\mathrm{b}}(t) + P_{\mathrm{sc}}(t) \geq \left|\left| \begin{matrix}
        2P_{\mathrm{ib}}(t) \\ P_{\mathrm{ib}}(t) - P_{\mathrm{b}}(t) - P_{\mathrm{sc}}(t)
    \end{matrix} \right|\right|_2,
\end{equation}
where $P_{\mathrm{ib}}(t)$ is the internal battery power and $P_{\mathrm{sc}}(t)$ is the short-circuit power, which we model as a piece-wise linear function of $E_{\mathrm{b}}(t)$ following \cite{ClementeSalazarEtAl2024, VerbruggenSalazarEtAl2019}. Subsequently, we calculate the energy stored in the BESS using
\begin{equation}
    \label{eq:euler_int_batt}
    \dot{E}_{\mathrm{b}}(t) = - P_{\mathrm{ib}}(t),
\end{equation}
where there are upper ($\overline{E}_{\mathrm{b}}$) and lower ($\underline{E}_{\mathrm{b}}$) limits for $E_\mathrm{b}(t)$ and, similarly, $P_{\mathrm{b}}(t)$ must be between an upper ($\overline{P}_{\mathrm{b}}$) and lower ($\underline{P}_{\mathrm{b}}$) bound.
Lastly, we use a periodic constraint on $E_\mathrm{b}(t)$ for the optimisation window of length $T$ (e.g., \unit[24]{h}):
\begin{equation}
\label{eq:continuity_batt}
    E_{\mathrm{b}}(0) = E_{\mathrm{b}}(T).
\end{equation}

We model battery degradation as a linear version of~\cite{BhandariSunEtAl2018,PaparellaLabeeEtAl2024}, where the amortisation of the battery per kWh
\begin{equation}
    \label{eq:C_amort_bat}
    a_{\mathrm{b}} = \frac{C_\mathrm{b, life}\cdot E_{\mathrm{b}}^\mathrm{inst}}{l_\mathrm{b} \cdot E_\mathrm{b}^\mathrm{cyc}},
\end{equation}
where $C_\mathrm{b, life}$ is the lifetime cost of the battery per installed kWh,
$E_{\mathrm{b}}^\mathrm{inst}$ is the installed battery capacity in kWh, the battery lifetime $l_\mathrm{b}$ is  the total number of cycles and $E_\mathrm{b}^\mathrm{cyc}$ is the amount of energy charged and discharged per cycle. We calculate the BESS amortisation costs with the net internal usage of the battery as a relaxed linear constraint
\begin{equation}
    \label{eq:opt_C_amort_BESS_relax}
    c_{\mathrm{b}}(t) \geq \left\{ \begin{matrix}
        a_{\mathrm{b}} \cdot P_{\mathrm{ib}}(t)\\
        a_{\mathrm{b}} \cdot -P_{\mathrm{ib}}(t)
    \end{matrix} \right.\,,
\end{equation}
where $c_{\mathrm{b}}(t)$ is the instantaneous amortisation cost rate at time $t$ for the BESS. 

\subsection{Power Grid}
The charging station can exchange power with the grid, incurring a variable cost
\begin{equation*} \label{eq:buying_constraint_unrelaxed}
    c_{\mathrm{el}}(t) = \left\{ \begin{matrix}
        p_{\mathrm{imp}}(t) \cdot P_{\mathrm{g}}(t) & \text{if } P_{\mathrm{g}}(t)\geq 0 \\
        p_{\mathrm{exp}}(t) \cdot P_{\mathrm{g}}(t) & \text{otherwise}
    \end{matrix} \right.\,,
\end{equation*}
where $p_{\mathrm{imp}}(t)$ and $p_{\mathrm{exp}}(t)$ are the importing and exporting prices at time $t$ respectively. We assume that $p_{\mathrm{imp}}(t) \geq p_{\mathrm{exp}}(t)$ and given that we aim to minimise the total energy cost, we can losslessly relax this equation to 
\begin{equation}
    \label{eq:buying_constraint}
    c_{\mathrm{el}}(t) \geq \left\{ \begin{matrix}
        p_{\mathrm{imp}}(t) \cdot P_{\mathrm{g}}(t)\\
        p_{\mathrm{exp}}(t) \cdot P_{\mathrm{g}}(t)
    \end{matrix} \right.\,,
\end{equation}
where $P_\mathrm{g}(t)$ is bounded by the energy contract or grid infrastructure limitation as
\begin{equation}
	\label{eq:grid_limits}
	P_\mathrm{g}(t) \in [\underline{P}_\mathrm{g}, \overline{P}_\mathrm{g} ]
\end{equation}

In addition, we consider the costs of the grid infrastructure. Since these costs depend on the size of the grid connection and thus the peak amount of concurrent EV charging sessions, we calculate the amortisation costs depending on the energy charged in each charging session
\begin{equation}
    \label{eq:grid_amort_price}
    a_\mathrm{g} = \frac{C_\mathrm{g, life} \cdot \overline{P}_\mathrm{g}}{E_\mathrm{ch, l_g}},
\end{equation}
where $a_\mathrm{g}$ is the average amortisation of the grid per kWh charged to an EV, $C_\mathrm{g, life}$ is the lifetime cost for the grid connection per kW of installed capacity 
and $E_\mathrm{ch, l_g}$ is the expected total amount of energy charged to vehicles in the lifetime of the grid infrastructure.

\subsection{Solar Plant}
The amortisation costs associated with the installation and maintenance of the PV 
panels are constant over time as
\begin{equation}
    \label{eq:C_amort_PV}
    a_{\mathrm{pv}} = \frac{C_\mathrm{pv, life}\cdot P_{\mathrm{pv}}^\mathrm{inst}}{l_\mathrm{pv}},
\end{equation}
where $a_{\mathrm{pv}}$ is the amortisation cost of the PV installation per time unit, $C_\mathrm{pv, life}$ is the lifetime cost for the PV panels per kW of installed peak generation power, $l_\mathrm{pv}$ is the lifetime of the PV panels and $P_{\mathrm{pv}}^\mathrm{inst}$ is the peak generation power of the PV installation.

\subsection{EV charging strategies}
\label{ch:EV charging strategies}
We model the charging process assuming perfect knowledge for each user $n$ of the energy desired $E_{\mathrm{ch},n}$ and the arrival and departure times of the session. We introduce a new time variable  $\tau_n = t - t_{\mathrm{c}, n}$
which tracks time for user $n$ relative to the origin set by the connection time $t_{\mathrm{c}, n}$. We model the charging process linearly as
\begin{equation}
	\label{eq:euler_EV_Es}
	\dot{E}_{\mathrm{ev}, n}(\tau_n) = P_{\mathrm{ch},n}(\tau_n),
\end{equation}
where the charging power $P_{\mathrm{ch},n}(\tau_n)$ can modify the energy level $E_{\mathrm{ev},n}(\tau_n)$ until disconnection time $\tau_{\mathrm{d},n}$ when $E_{\mathrm{ev},n}(\tau_n=\tau_{\mathrm{d},n}) = E_{\mathrm{ch},n}$.

We consider three EV charging strategies in this work. The simplest charging strategy is nominal charging, in which EVs start charging at full power until the energy request is fulfilled. This is formulated as
\begin{equation}
\label{eq:Nominal_strat}
    P_{\mathrm{ch},n}(\tau_n) = \left\{ \begin{aligned}
        \overline{P}_{\mathrm{ch}} & ~\text{if } 0 \leq \tau_n \leq \tau_{\mathrm{ch}, n}\\
        0 & ~\text{if } \tau_n > \tau_{\mathrm{ch}, n}
    \end{aligned} \right.,
\end{equation}
where the time needed to meet the demand is $\tau_{\mathrm{ch}, n} = \frac{E_{\mathrm{ch},n}}{\overline{P}_{\mathrm{ch}}}$. Then, we consider two smart strategies. First, V2G allows charging to dynamically adjust within its limits as
\begin{equation}
	\label{eq:smart_strat}
	P_{\mathrm{ch},n}(\tau_n) \in  \left[  \underline{P}_{\mathrm{ch}}, \overline{P}_{\mathrm{ch}} \right]. 
\end{equation}
 Second, V1G is a variant in which we prevent the EV battery from discharging as
 \begin{equation}
 	\label{eq:V1G_strat}
 	\underline{P}_{\mathrm{ch}}=0.
 \end{equation}
 We can obtain the total EV power demand at time $t$ as
\begin{equation}
	\label{eq:sum_Pev}
	P_{\mathrm{ev}}(t) = \sum_{n=1}^{N_{\mathrm{ev}}} P_{\mathrm{ch},n}(t - t_{\mathrm{c}, n})\cdot \mathds{1}_{\tau_n\geq0},
\end{equation}
where $N_\mathrm{ev}$ is the total number of users in the optimisation window and $\mathds{1}_x$ is an indicator function set to 1 if $x$ is true and 0 otherwise.

We model EV battery degradation similarly to~\cref{eq:C_amort_bat,eq:opt_C_amort_BESS_relax}  to obtain the instantaneous amortisation cost rate $c_{\mathrm{evb},n}$ of the EV battery of user $n$.
Furthermore, we model the charger amortisation cost following~\cref{eq:grid_amort_price} as
\begin{equation}
    \label{eq:chargers_amort_price}
    a_\mathrm{ch} = \frac{C_\mathrm{ch, life} \cdot N_\mathrm{ch}}{E_{\mathrm{ch, l_{ch}}}},
\end{equation}
where $a_\mathrm{ch}$ is the amortisation cost per kWh charged to an EV for the chargers, $C_\mathrm{ch, life}$ are the lifetime costs for a single charger
, $N_\mathrm{ch}$ is the number of chargers at the energy hub and $E_{\mathrm{ch, l_{ch}}}$ is the expected total amount of energy charged to vehicles in the lifetime of the chargers.
\section{Optimal Hub Operation}
\label{ch:Optimal_hub_operation}

We operate the energy hub with the goal of minimising total operational cost. In this work, we consider a combination of two types of costs: economic costs $C^\mathrm{eco}$ and greenhouse gas emission costs $C^\mathrm{GHG}$. We can express the lifetime costs required to compute the corresponding amortisation and degradation costs from the previous section as
\begin{equation}
    \label{eq:relative_cost_alpha}
    C_\mathrm{life} = (1-\alpha)\cdot C^{\mathrm{eco}}_\mathrm{life} + \alpha\cdot C^{\mathrm{GHG}}_\mathrm{life},
\end{equation}
where $\alpha \in [0,1]$ indicates the weight of the function. Similarly, we compute importing and exporting prices in~\eqref{eq:buying_constraint} based on both metrics. This approach allows us to obtain a Pareto front by solving the optimisation problem for varying weights. 

The objective function $J$ in Problem~\ref{prob:simple_optimisation} considers both operator and user terms.  
We define the cost of the operator as
\begin{equation*}
	J_\mathrm{op} = \int_{0}^{T}
	c_{\mathrm{b}}(t) + c_{\mathrm{el}}(t)
	+ a_{\mathrm{pv}}\,\mathrm{d}t 
	 + \sum_{n=1}^{N_{\mathrm{ev}}}
	\bigl(a_{\mathrm{g}} + a_{\mathrm{ch}}\bigr) \cdot E_{\mathrm{ch},n}, 
\end{equation*}
which includes all the component costs covered by the energy hub operator. We can also define the cost of each user $n$ as
\begin{equation*}
	J_{\mathrm{ev}, n} =  \int_{0}^{\tau_{\mathrm{d},n}}
	c_{\mathrm{evb},n}(\tau_n)\,\mathrm{d}\tau_n.
\end{equation*}
Additionally, to ensure charging powers are evenly distributed between EVs over time, we include a regularisation term in $J$ as 
\begin{equation*}
    J =  J_\mathrm{op} + \sum_{n=1}^{N_\mathrm{ev}} J_{\mathrm{ev},n} + \alpha_\mathrm{r} \cdot \int_{0}^{\tau_{\mathrm{d}, n}}||P_{\mathrm{ch},n}(\tau_n)||^2_2 ~\mathrm{d}\tau_n,
\end{equation*}
where $\alpha_\mathrm{r}$ is a regularisation weight.
We then formulate the optimal operation of the hub as Problem~\ref{prob:simple_optimisation}, 
where the decision variables are $P_{\mathrm{ch},n}$ and  $P_\mathrm{ib}$.

\begin{prob}[Optimal Hub Operation Problem]\label{prob:simple_optimisation}
\begin{equation*}
\begin{aligned}
    &\!\min_{P_{\mathrm{ch},n}, P_{\mathrm{ib}}} & & J \\
    & \textnormal{s.t. } & & \text{\cref{eq:Hub_Pwrbalance,eq:battery_loss_ib,eq:euler_int_batt,eq:continuity_batt},\cref{eq:buying_constraint,eq:opt_C_amort_BESS_relax,eq:grid_limits}, \cref{eq:euler_EV_Es}, \cref{eq:sum_Pev} and}\\
    & & & \begin{aligned}
        \cref{eq:Nominal_strat} &\text{ if Nominal} \\
        \cref{eq:smart_strat} &\text{ if V2G} \\
        \cref{eq:smart_strat},\cref{eq:V1G_strat} &\text{ if V1G}.
    \end{aligned}
\end{aligned}
\end{equation*}
\end{prob}

For comparison, we introduce a no-BESS configuration to the three charging strategies used in Problem~\ref{prob:simple_optimisation}, where we remove the BESS unit and employ the nominal charging strategy~\eqref{eq:Nominal_strat}.

\section{Profit sharing methods}
\label{ch:Game theory}
The smart charging strategies used to obtain optimal hub operation produce cost savings. This opens up the discussion on how to fairly distribute cost savings among participating users and charging station operators. 
In this section we propose a justice-informed cost allocation method based on Nash bargaining.
Furthermore, we present a sufficientarian approach to reduce the inequality caused by users' temporal inflexibilities.

\subsection{Nash Bargaining}
\label{ch:Nash_Bargaining}
We formulate this profit sharing problem as a Nash bargaining problem, which is part of cooperative bargaining problems. When agents generate a surplus by cooperating, Nash bargaining provides a trade-off among all agents, maximising agent improvement with respect to a fallback position.
In this work, we define the bargaining problem between EV users and the energy hub operator. The utility function of user $n$ is
\begin{equation}
	\label{eq:utility_ev}
	u_{\mathrm{ev},n} = - J_{\mathrm{ev}, n} - C_{\mathrm{ev}, n}^\mathrm{Nash},
\end{equation}
where $C_{\mathrm{ev},n}^\mathrm{Nash}$ is the charging cost paid by user $n$ under Nash bargaining. Similarly, the utility of the hub operator is
\begin{equation}
	\label{eq:utility_operator}
	u_\mathrm{op} = - J_\mathrm{op} + \sum_{n=1}^{N_\mathrm{ev}} C_{\mathrm{ev},n}^\mathrm{Nash}.
\end{equation}
In Nash bargaining, resource allocation maximises the product of utility improvements. We define the objective function of this maximisation problem as
\begin{equation}
    \label{eq:NBS1}
    \widehat{J}_\mathrm{Nash} = (u_\mathrm{op} - u_\mathrm{op}^\mathrm{d}) \cdot \prod_{n=1}^{N_\mathrm{ev}}\left(u_{\mathrm{ev}, n} - u_{\mathrm{ev}, n}^\mathrm{d}\right),
\end{equation}
where $u_\mathrm{op}^\mathrm{d}$ and $u_{\mathrm{ev}, n}^\mathrm{d}$ are the disagreement utilities of the operator and user $n$, respectively. We assume that in case of disagreement, the operator does not meet any charging demand and therefore has utility
\begin{equation}
	\label{eq:disagreement_operator}
	u_\mathrm{op}^\mathrm{d} = -J_\mathrm{op}^\mathrm{d},
\end{equation}
where $J^\mathrm{d}_\mathrm{op}$ is the optimal value of Problem~\ref{prob:simple_optimisation} with no charging demand. 
 For users, we assume charging takes place at a different station with utility
\begin{equation}
	u_{\mathrm{ev}, n}^\mathrm{d} = -J_{\mathrm{ev},n}^\mathrm{d} - C_{\mathrm{ev},n}^\mathrm{d},
\end{equation}
where  $J_{\mathrm{ev},n}^\mathrm{d}$ is the battery degradation cost of user $n$ under nominal charging in~\eqref{eq:Nominal_strat} and the disagreement cost $C_{\mathrm{ev}, n}^{\mathrm{d}}$ is
\begin{multline}
	\label{eq:disagreement_nominal}
	C_{\mathrm{ev}, n}^{\mathrm{d}} =  \beta \cdot \int_{0}^{\tau_{\mathrm{d}, n}} p_{\mathrm{imp}}(\tau_n + t_{\mathrm{c},n}) \cdot P_{\mathrm{ch}, n}^{\mathrm{nom}}(\tau_n) ~\mathrm{d}\tau_n \\
	+ \beta \cdot \left( a_\mathrm{g} + a_\mathrm{ch} \right) \cdot E_{\mathrm{ch}, n},
\end{multline}
where $\beta>1$ is multiplier considering the operator profit margin and $P_{\mathrm{ch}, n}^{\mathrm{nom}}$ is the nominal charging strategy in~\cref{eq:Nominal_strat}. 

We follow \cite{Kennan2025} and use a log-transform for the non-linear objective function in~\eqref{eq:NBS1}. 
Furthermore, we introduce a weight so that users get a utility improvement proportional to the energy charged. The transformed utilities of users $ v_{\mathrm{ev}, n} $ and the operator $v_{\mathrm{op}}$  are
\begin{equation}
    \label{eq:log_op_weight}
    v_{\mathrm{op}} = \sum_{n=1}^{N_\mathrm{ev}}E_{\mathrm{ch}, n} \cdot \log \left(\frac{u_\mathrm{op}-u_\mathrm{op}^{\mathrm{d}}}{\sum_{n=1}^{N_\mathrm{ev}}E_{\mathrm{ch}, n}} \right),
\end{equation}
\begin{equation}
    \label{eq:log_ev_weight}
    v_{\mathrm{ev}, n} = E_{\mathrm{ch}, n} \cdot \log\left(\frac{u_{\mathrm{ev}, n} - u_{\mathrm{ev}, n}^\mathrm{d}}{E_{\mathrm{ch}, n}}\right),
\end{equation}
which can be implemented using an exponential cone resulting in an exponential cone optimisation problem. We constrain the utility improvements by
\begin{equation}
	\label{eq:constr_util_op}
	u_\mathrm{op}-u_\mathrm{op}^{\mathrm{d}} \geq 0,
\end{equation}
\begin{equation}
	\label{eq:constr_util_ev}
	u_{\mathrm{ev}, n} - u_{\mathrm{ev}, n}^\mathrm{d} \geq 0 \quad \forall~ n \in \{1, ..., N_\mathrm{ev}\}.
\end{equation}
The equivalent minimisation objective function of the problem is the negative of the transformation of~\eqref{eq:NBS1} 
\begin{equation*}
	J_\mathrm{Nash} = - v_{\mathrm{op}} - \sum_{n=1}^{N_\mathrm{ev}}v_{\mathrm{ev}, n}.
\end{equation*}

The resulting Nash bargaining optimisation problem is then as follows.
\begin{prob}[Nash bargaining problem]\label[prob]{prob:NBS}
\begin{equation*}
\begin{aligned}
    &\!\min_{P_{\mathrm{ch},n}, P_{\mathrm{ib}}, C_{\mathrm{ev},n}^\mathrm{Nash}} & & J_\mathrm{Nash} \\
    & \textnormal{s.t. } & & \text{\cref{eq:Hub_Pwrbalance,eq:battery_loss_ib,eq:euler_int_batt,eq:continuity_batt,eq:buying_constraint,eq:opt_C_amort_BESS_relax,eq:grid_limits,eq:euler_EV_Es,eq:smart_strat,eq:sum_Pev}}\\
    &&&
    \text{\cref{eq:utility_ev,eq:utility_operator,eq:constr_util_op,eq:constr_util_ev,eq:log_op_weight,eq:log_ev_weight}}
\end{aligned}
\end{equation*}
\end{prob}

\subsection{Sufficientarian Nash Bargaining}
Nash bargaining allocates savings by treating all agents as symmetric strategic actors, maximising the product of utility improvements relative to a disagreement point. This symmetry, however, does not account for the fact that users differ in how much flexibility they have to adapt their charging behaviour. 
EV charging sessions must be planned around a user’s schedule in a way that conventional refuelling does not, placing a disproportionate burden on users with tight schedules or no access to home charging. As shown in the literature on flexibility capital~\cite{Libertson2022}, lower-income households tend to have less scheduling flexibility, leaving them unable to shift charging to cheaper time windows~\cite{MalabananSauriLavieriEtAl2025}. Under plain Nash bargaining, these users tend to pay higher prices for constraints they cannot control, compounding existing inequalities~\cite{ZhuLiuEtAl2026}.

To mitigate this, we introduce a sufficientarian extension of Nash bargaining~\cite{SalazarBetancourGiraldoEtAl2024,SalazarBetancurGiraldoEtAl2025}. 
Sufficientarianism holds that the primary distributive objective is not to equalise outcomes, but to ensure that everyone reaches a sufficient threshold of well-being or access to essential opportunities ~\cite{Nussbaum2011,Martens2017}. In our context, this means ensuring that all users have access to charging at an affordable price, irrespective of their capacity to shift demand.
We implement this by adding an upper limit $\overline{p}_\mathrm{ev}$ on the per-kWh charging price, ensuring that the most constrained users are not charged beyond a threshold consistent with affordable household energy expenditure. We represent this limit by adding another constraint to Problem~\ref{prob:NBS}:
\begin{equation}
    C_{\mathrm{ev}, n}^\mathrm{Nash}/E_{\mathrm{ch}, n} \leq \overline{p}_\mathrm{ev},
\end{equation}
where $\overline{p}_\mathrm{ev}$ is the upper limit on the charging price.
We set $\overline{p}_\mathrm{ev}$ at the 90th percentile of the Nash bargaining price distribution, using a quantile to locate the threshold within the price distribution~\cite{HyndmanFan1996}. The constraint operationalises the sufficientarian objective of protecting users from excessive charging costs. However, sufficientarianism does not prescribe a specific price threshold, making its calibration context-dependent. The choice should therefore be examined through sensitivity analysis of alternative thresholds, considering the trade-off between protecting financially constrained users and maintaining an efficient allocation.

\section{Case Study}
\label{ch:Case_study}
\label{ch:Case2_Caltech}
For this case study, we consider a charging energy hub with adapted data from different sources. The EV charging demand dataset belongs to a university campus in California, US, using Level 2 AC chargers~\cite{LeeLiEtAl2019}. We size the components of the hub according to the total charging demand in the hub. We use hourly day-ahead electricity price and emission information from the Netherlands for 2025 \cite{Ember2026, NED2026}. On top of this, we add the importing electricity tax to the economic cost for the Netherlands \cite{Belastingdienst2026}. For the PV power, we use data from a solar farm in California, US, \cite{OpenEI2026}. Based on the data available, we work in discrete time steps with a resolution $\Delta t = \unit[15]{min}$ and use the forward Euler method to discretise~\cref{eq:euler_int_batt,eq:euler_EV_Es}.
We use an optimisation window of 5 days.
\ifextendedversion
\autoref{tab:Optimisation_parameters_fairness}
\else
The extended version of this paper~\cite{arxiv?}
\fi
\noindent shows additional parameters used in this case study.

To compute the amortisation costs of the components, we determine their lifetime costs  expressed in economic terms and as CO$_2$ equivalent emissions. We compute economic component lifetime cost as $C_\mathrm{life}^\mathrm{eco} = \mathrm{CAPEX} + \mathrm{OPEX} \cdot l$, considering the capital expenditure (CAPEX) and the operational expenditure (OPEX) of the component over its lifetime $l$. For the component lifetime cost expressed in GHG emissions $C_\mathrm{life}^\mathrm{GHG}$,  we use the global warming potential value stemming from component lifetime analysis.
For the EV batteries, we assume identical amortisation parameters across users. Since we do not know the battery type and capacity of the EVs, we assume that during a cycle the battery is fully charged and discharged. 
\ifextendedversion
\autoref{tab:Cost_parameters_fairness}
\else
The extended version of this paper~\cite{arxiv?}
\fi
\noindent shows the values used to determine all amortisation costs in this case study.

\subsection{Pareto Front in Optimal Hub Operation}
\label{ch:results_case_studies}
\begin{figure}[t]
	\centering
	\includegraphics[width=0.9\linewidth]{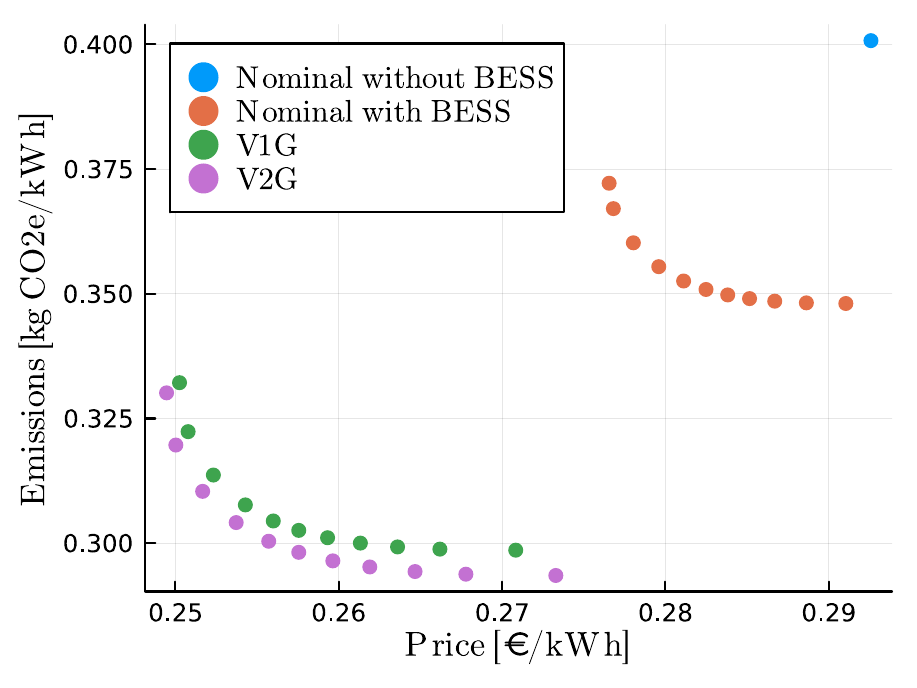}
	\caption{Pareto fronts for the cost per kWh charged to an EV at the charging hub for various charging strategies, averaged over all vehicles for multiple instances in a year for the case study.}
	\label{fig:strategies_alphas_Caltech}
\end{figure}

In this section, we report the results of optimal hub operation for different objective functions, based on a combination of economic and emission costs. We vary the weight $\alpha$ in~\eqref{eq:relative_cost_alpha} and solve Problem~\ref{prob:simple_optimisation} for 24 days in the data-set selected to cover all months of the year. We obtain the Pareto front for all three charging strategies in Fig.~\ref{fig:strategies_alphas_Caltech}, where each point is the average operation and component-amortisation cost, relative to the total EV energy demand in the optimisation window. Additionally, we include the result for the hub configuration without a BESS unit and using nominal charging.

Given that renewable energy generation is cheaper than fuel-based energy generation \cite{IEA2025}, we expect economic-optimal operation also produces emission-optimal operation. However, for varying values of $\alpha$ in Fig.~\ref{fig:strategies_alphas_Caltech} there is a trade-off between minimising economic and emission costs. On top of this, we see that increasing flexibility allows to obtain lower average economic and emission costs for all values of $\alpha$. The nominal strategy without BESS has the lowest flexibility and, consequently, the highest average economic and emission cost. By adding the BESS, energy hub operation improves. Significant cost and emission savings are possible when enabling smart charging.  When selecting the V1G charging strategy, EV charging can be shifted to more convenient times.
Furthermore, V2G charging yields a marginal improvement with respect to V1G, thanks to the additional flexibility provided by a small number of users. A larger V2G improvement would require users to stay connected for longer periods of time, allowing the controller to charge and discharge EVs during horizons with larger price variations and, consequently, to offset the increase in EV battery degradation costs.

\subsection{Fair Cost Distribution}
\label{Ch:Case_study_fairness}
In this section, we compare the cost distribution under the Nash bargaining solution in Problem~\ref{prob:NBS} with that of its sufficientarian alternative. The latter introduces a sufficientarian concern for protecting users from excessively burdensome charging costs, particularly those with limited financial resources and scheduling flexibility. We set  $\overline{p}_\mathrm{ev}$ at the 90th percentile of the Nash bargaining price distribution, yielding \texteuro 0.369/kWh in this case study. This imposes an upper limit on charging prices, constraining the highest prices in the distribution while retaining scope for price differentiation among users.

\begin{figure}[t]
	\centering
	\includegraphics[width=0.9\linewidth]{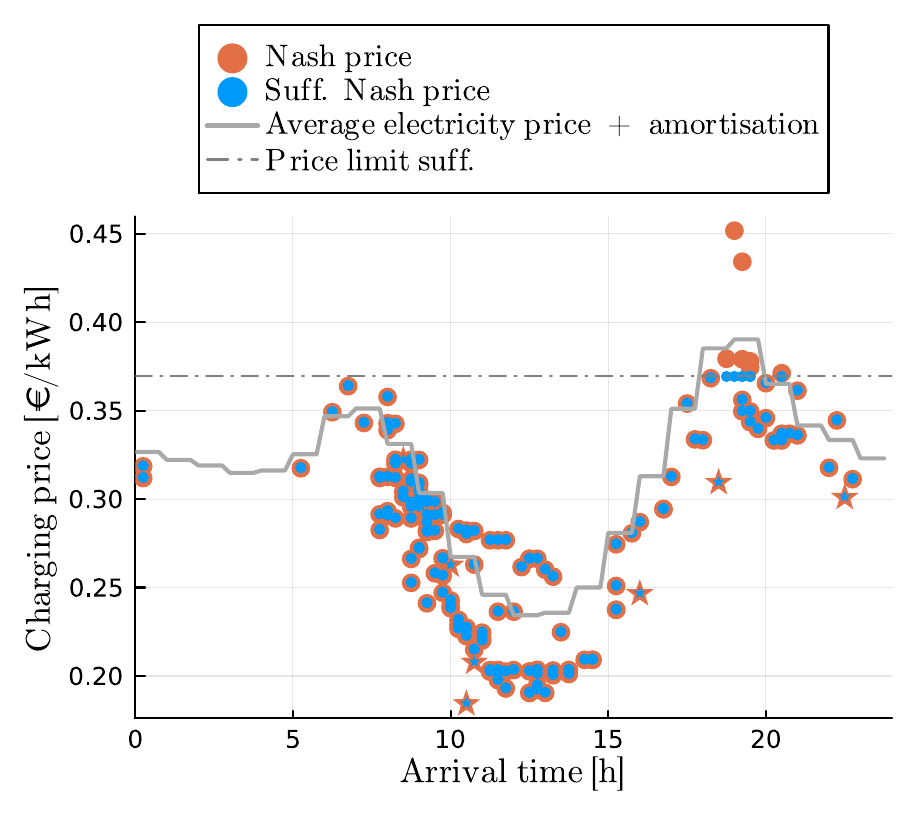}
	\caption{Charging prices for sessions during a sample \unit[120]{h} optimisation window using different cost distribution principles. V1G charging is indicated with a circle, V2G charging with a star. We include the 5-day average hourly grid electricity price and charger amortisation costs.}
	\label{fig:Nash_price_starttime}
\end{figure}

Fig.~\ref{fig:Nash_price_starttime} presents the charging price $\frac{C_{\mathrm{ev},n}^\mathrm{Nash}}{E_{\mathrm{ch},n}}$, after distributing costs through the Nash bargaining method, at a sample day with a \unit[120]{h} optimisation horizon. For all V1G users, these prices come from the same price reduction relative to the disagreement price $\frac{C_{\mathrm{ev},n}^\mathrm{d}}{E_{\mathrm{ch},n}}$. V2G users enjoy a greater price reduction, which compensates for their increased battery degradation. 
We also include the sufficientarian Nash bargaining price and the upper limit price in Fig.~\ref{fig:Nash_price_starttime}. This approach successfully prevents disadvantaged users from paying over the limit, while only slightly affecting the prices of others. We can also see that even with a sufficientarian cost distribution, we can see mid-day arriving users pay significantly less than the rest, highlighting the influence of the disagreement price in our approach.

\begin{figure}
    \centering
    \includegraphics[width=0.9\linewidth]{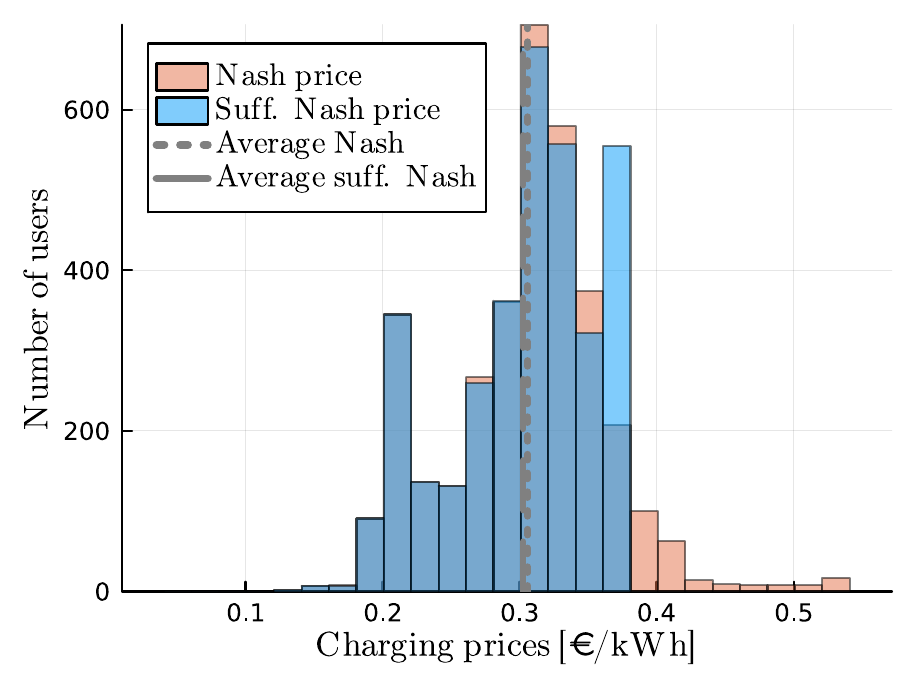}
    \caption{Histograms of the charging prices throughout a year for the Nash and sufficientarian Nash distributions for the case study.}
    \label{fig:Histograms_Nash-S_Nash_campus}
\end{figure}

We extend the comparison of the two methods over a longer time horizon by solving the cost distribution problem for 24 days across the year, selecting two days per month and considering a \unit[120]{h} horizon window. Table~\ref{tab:prices_averaged} reports the average charging price and daily station profit across the year. Both distribution methods substantially improve upon the disagreement outcome. Table~\ref{tab:prices_averaged} shows that average charging prices under Nash bargaining and the sufficientarian alternative differ by only 1\%, while their average station profits differ by 5\%. Despite these relatively small differences in aggregate outcomes, Fig.~\ref{fig:Histograms_Nash-S_Nash_campus} reveals substantial differences in how charging costs are distributed among users under the two methods. The sufficientarian Nash bargaining approach respects the upper price limit by shifting the price distribution towards this threshold, thereby improving affordability for users with low flexibility capital. As a result, although the two methods yield comparable aggregate outcomes, their distributional effects differ considerably at the individual level. The choice of profit-sharing method can therefore have markedly different implications for individual users, depending on their flexibility capital: Still promoting flexibility when it is possible, whilst not punishing when it is not.

\begin{table}[tbp]
	\centering
	\caption{The charging prices and daily EVCS profit for all considered allocation methods averaged over the year.}
	\label{tab:prices_averaged}
	\begin{tabular}{l|c|c}
		Allocation Method & Charging price & EVCS profit \\
		& [\texteuro/kWh] & [\texteuro/day] \\ \hline
		Disagreement Outcome        & 0.348 & 6.94  \\
		Nash Bargaining      & 0.309 & 17.09 \\
		Sufficientarian Nash & 0.306 & 16.28 \\
	\end{tabular}
\end{table}

\section{Conclusions} \label{ch:conclusion}
In this paper, we introduced multi-objective optimal charging strategies and profit sharing methods for an electric vehicle (EV) charging energy hub equipped with a battery storage system, solar panels and a grid connection.
Our case study results showed how a stationary battery and dynamic vehicle charging operations, including V2G, can significantly reduce total costs and emissions. However, these savings are only possible if there is cooperation between the operator and users.
The second part of this paper examined how the savings generated through such a cooperation can be distributed between EV users and the charging station operator. We compared Nash bargaining with a sufficientarian extension that limits maximum charging prices to protect users with low flexibility capital. This constraint improves the fairness of the cost distribution while having little impact on V2G incentives, average charging prices and station profits.
Ultimately, it shows that accounting for inequalities in flexibility capital can make EV charging more equitable without substantially compromising its economic benefits.

This work opens the field for the following possible extensions. Currently, the cost allocation methods solely take the economic cost into account. Future work should also consider the emissions as they also impact social welfare~\cite{PolancoVasquezLopezRedondoEtAl2023,XieChen2024}.
Furthermore, we plan to implement the methods and charging strategies in real time.
Finally, future research should investigate how the sufficientarian price threshold can be normatively grounded and systematically calibrated, taking into account differences in household income, charging needs, and scheduling flexibility.

\section*{Acknowledgment}
This work was supported by RVO within the Charging Energy Hubs project with number NGFS23020 and was partially funded by the Dutch National Growth Fund.

\bibliographystyle{ieeetr}

\bibliography{energyhubs.bib, SML_papers.bib, mobility.bib} 

\newcommand{\noopsort}[1]{} \newcommand{\printfirst}[2]{#1}
  \newcommand{\singleletter}[1]{#1} \newcommand{\switchargs}[2]{#2#1}
\begin{thebibliography}{10}

\bibitem{IEA2024b}
IEA, ``Renewables 2024,'' tech. rep., {International Energy Agency}, 2024.

\bibitem{MliloBrownEtAl2021}
N.~Mlilo, J.~Brown, and T.~Ahfock, ``Impact of intermittent renewable energy
  generation penetration on the power system networks – a review,'' {\em
  Technology and Economics of Smart Grids and Sustainable Energy}, vol.~6,
  no.~1, 2021.

\bibitem{AnwarMuratoriEtAl2022}
M.~B. Anwar, M.~Muratori, P.~Jadun, E.~Hale, B.~Bush, P.~Denholm, O.~Ma, and
  K.~Podkaminer, ``Assessing the value of electric vehicle managed charging: a
  review of methodologies and results,'' {\em Energy \& Environmental Science},
  vol.~15, no.~2, pp.~466--498, 2022.

\bibitem{GeSimeoneEtAl2021}
Y.~Ge, C.~Simeone, A.~Duvall, and E.~Wood, ``There’s no place like home:
  Residential parking, electrical access, and implications for the future of
  electric vehicle charging infrastructure,'' tech. rep., {National Renewable
  Energy Lab (NREL)}, 2021.

\bibitem{Libertson2022}
F.~Libertson, ``(no) room for time-shifting energy use: Reviewing and
  reconceptualizing flexibility capital,'' {\em Energy Research \& Social
  Science}, vol.~94, p.~102886, 2022.

\bibitem{MalabananSauriLavieriEtAl2025}
I.~Malabanan, P.~Sauri~Lavieri, and I.~Mateo-Babiano, ``Electric vehicle
  charging disadvantage: a social justice perspective on charging with
  implications to accessibility,'' {\em Transport Reviews}, vol.~45, no.~5,
  pp.~696--725, 2025.

\bibitem{GhanbariMotlaghOladigboluEtAl2025}
S.~Ghanbari~Motlagh, J.~Oladigbolu, and L.~Li, ``A review on electric vehicle
  charging station operation considering market dynamics and grid
  interaction,'' {\em {Applied Energy}}, vol.~392, p.~126058, 2025.

\bibitem{LeFlochBansalEtAl2019}
C.~Le~Floch, S.~Bansal, C.~J. Tomlin, S.~J. Moura, and M.~N. Zeilinger,
  ``Plug-and-play model predictive control for load shaping and voltage control
  in smart grids,'' {\em {IEEE Transactions on Smart Grid}}, vol.~10, no.~3,
  pp.~2334--2344, 2019.

\bibitem{PolancoVasquezLopezRedondoEtAl2023}
L.~O. Polanco~Vásquez, J.~López~Redondo, J.~D. Álvarez Hervás, V.~M.
  Ramírez, and J.~L. Torres, ``Balancing {CO2} emissions and economic cost in
  a microgrid through an energy management system using mpc and multi-objective
  optimization,'' {\em {Applied Energy}}, vol.~347, p.~120998, 2023.

\bibitem{CeustersRodriguezEtAl2021}
G.~Ceusters, R.~C. Rodríguez, A.~B. García, R.~Franke, G.~Deconinck,
  L.~Helsen, A.~Nowé, M.~Messagie, and L.~R. Camargo, ``Model-predictive
  control and reinforcement learning in multi-energy system case studies,''
  {\em {Applied Energy}}, vol.~303, p.~117634, 2021.

\bibitem{FernandezZapicoHofmanEtAl2025}
D.~Fernandez~Zapico, T.~Hofman, and M.~Salazar, ``Stochastic model predictive
  control of charging energy hubs with conformal prediction,'' in {\em {Proc.\
  IEEE Conf.\ on Decision and Control}}, 2025.

\bibitem{PatilKalkhambkar2020}
H.~Patil and V.~N. Kalkhambkar, ``Charging cost minimisation by centralised
  controlled charging of electric vehicles,'' {\em {International Transactions
  on Electrical Energy Systems}}, vol.~30, no.~2, p.~e12226, 2020.

\bibitem{ZhangHoreshEtAl2023}
R.~Zhang, N.~Horesh, E.~Kontou, and Y.~Zhou, ``Electric vehicle community
  charging hubs in multi-unit dwellings: Scheduling and techno-economic
  assessment,'' {\em {Transportation Research Part D: Transport and
  Environment}}, vol.~120, p.~103776, 2023.

\bibitem{GanTopcuEtAl2013}
L.~Gan, U.~Topcu, and S.~H. Low, ``Optimal decentralized protocol for electric
  vehicle charging,'' {\em {IEEE Transactions on Power Systems}}, vol.~28,
  no.~2, pp.~940--951, 2013.

\bibitem{AounAddaEtAl2024}
A.~Aoun, M.~Adda, A.~Ilinca, M.~Ghandour, and H.~Ibrahim, ``Dynamic charging
  optimization algorithm for electric vehicles to mitigate grid power peaks,''
  {\em {World Electric Vehicle Journal}}, vol.~15, no.~7, p.~324, 2024.

\bibitem{KarmakerHossainEtAl2023}
A.~K. Karmaker, M.~A. Hossain, H.~R. Pota, A.~Onen, and J.~Jung, ``Energy
  management system for hybrid renewable energy-based electric vehicle charging
  station,'' {\em {IEEE Access}}, vol.~11, pp.~27793--27805, 2023.

\bibitem{HoehneChester2016}
C.~G. Hoehne and M.~V. Chester, ``Optimizing plug-in electric vehicle and
  vehicle-to-grid charge scheduling to minimize carbon emissions,'' {\em
  {Energy}}, vol.~115, pp.~646--657, 2016.

\bibitem{LeeMadanat2017}
J.~Lee and S.~Madanat, ``Optimal design of electric vehicle public charging
  system in an urban network for greenhouse gas emission and cost
  minimization,'' {\em {Transportation Research Part C: Emerging
  Technologies}}, vol.~85, pp.~494--508, 2017.

\bibitem{AldosariAliEtAl2024}
O.~Aldosari, Z.~M. Ali, S.~H.~E. Abdel~Aleem, and M.~H. Mostafa, ``Optimizing
  microgrid performance: Strategic integration of electric vehicle charging
  with renewable energy and storage systems for total operation cost and
  emissions minimization,'' {\em {PLoS ONE}}, vol.~19, no.~10, p.~e0307810,
  2024.

\bibitem{FrolikHines2012}
J.~Frolik and P.~Hines, ``Urgency-driven, plug-in electric vehicle charging,''
  in {\em 2012 3rd {IEEE} {PES} Innovative Smart Grid Technologies Europe
  {(ISGT Europe)}}, 2012.

\bibitem{PaudelWolfEtAl2024}
D.~Paudel, L.~Wolf, and T.~K. Das, ``Joint operation of a fast-charging {EV}
  hub with a stand-alone independent battery storage system under fairness
  considerations.'' ArXiv, 2024.
\newblock \url{https://arxiv.org/abs/2404.16113}.

\bibitem{BahramiToulabiEtAl2018}
S.~Bahrami, M.~Toulabi, S.~Ranjbar, M.~Moeini-Aghtaie, and A.~M. Ranjbar, ``A
  decentralized energy management framework for energy hubs in dynamic pricing
  markets,'' {\em {IEEE Transactions on Smart Grid}}, vol.~9, no.~6,
  pp.~6780--6792, 2018.

\bibitem{XieChen2024}
R.~Xie and Y.~Chen, ``Real-time bidding strategy of energy storage in an energy
  market with carbon emission allocation based on {Aumann-Shapley} prices,''
  {\em {IEEE Transactions on Energy Markets, Policy and Regulation}}, vol.~2,
  no.~3, pp.~350--367, 2024.

\bibitem{LimmerDietrich2018}
S.~Limmer and M.~Dietrich, ``Optimization of dynamic prices for electric
  vehicle charging considering fairness,'' in {\em 2018 {IEEE} Symposium Series
  on Computational Intelligence {(SSCI)}}, 2018.

\bibitem{ShilovElokdaEtAl2026}
I.~Shilov, E.~Elokda, S.~Hall, H.~H. Nax, and S.~Bolognani, ``Welfare and cost
  aggregation for multi-agent control: When to choose which social cost
  function, and why?,'' {\em IEEE Open Journal of Control Systems}, vol.~5,
  pp.~80--90, 2026.

\bibitem{ClementeSalazarEtAl2024}
M.~Clemente, M.~Salazar, and T.~Hofman, ``Concurrent {D}esign {O}ptimization of
  {P}owertrain {C}omponent {M}odules in a {F}amily of {E}lectric {V}ehicles,''
  {\em {Applied Energy}}, vol.~379, p.~124840, 2025.

\bibitem{VerbruggenSalazarEtAl2019}
F.~J.~R. Verbruggen, M.~Salazar, M.~Pavone, and T.~Hofman, ``Joint design and
  control of electric vehicle propulsion systems,'' in {\em {European Control
  Conference}}, 2020.

\bibitem{BhandariSunEtAl2018}
V.~Bhandari, K.~Sun, and F.~Homans, ``The profitability of vehicle to grid for
  system participants - a case study from the electricity reliability council
  of {Texas},'' {\em {Energy}}, vol.~153, pp.~278--286, 2018.

\bibitem{PaparellaLabeeEtAl2024}
F.~Paparella, P.~Labee, S.~Wilkins, T.~Hofman, S.~Rasouli, and M.~Salazar,
  ``Multi-objective optimal trade-off between v2g activities and battery
  degradation in electric mobility-as-a-service systems,'' in {\em {Proc.\ IEEE
  Int.\ Conf.\ on Intelligent Transportation Systems}}, 2024.

\bibitem{Kennan2025}
J.~Kennan, ``The {Nash} bargaining solution with many players.'' unpublished,
  2025.

\bibitem{ZhuLiuEtAl2026}
X.~Zhu, S.~Liu, Y.~Cao, and Y.~Zhang, ``{An asymmetric Nash bargaining based
  day-ahead distributed energy sharing strategy for integrated energy service
  providers},'' {\em {International Journal of Electrical Power \& Energy
  Systems}}, vol.~169, p.~110744, 2025.

\bibitem{SalazarBetancourGiraldoEtAl2024}
M.~Salazar, S.~Betancur~Giraldo, F.~Paparella, and L.~Pedroso, ``On
  accessibility fairness in intermodal mobility-on-demand systems,'' in {\em
  {Proc.\ of the IFAC Symposium on Control in Transportation Systems}}, 2024.

\bibitem{SalazarBetancurGiraldoEtAl2025}
M.~Salazar, S.~Betancur~Giraldo, F.~Paparella, L.~Pedroso, and K.~Martens,
  ``Mobilizing transport justice: A sufficientarian optimization framework for
  intermodal mobility systems,'' {\em {NPJ Sustainable Mobility and
  Transport}}, vol.~2, no.~43, 2025.

\bibitem{Nussbaum2011}
M.~Nussbaum, {\em Creating Capabilities}.
\newblock Harvard University Press, 2011.

\bibitem{Martens2017}
K.~Martens, {\em Transport Justice -- Designing Fair Transportation Systems}.
\newblock {Taylor \& Francis}, 2017.

\bibitem{HyndmanFan1996}
R.~J. Hyndman and Y.~Fan, ``{Sample quantiles in statistical packages},'' {\em
  {The American Statistician}}, vol.~50, no.~4, pp.~361--365, 1996.

\bibitem{LeeLiEtAl2019}
Z.~J. Lee, T.~Li, and S.~H. Low, ``{ACN-Data:} analysis and applications of an
  open {EV} charging dataset,'' in {\em Proceedings of the Tenth International
  Conference on Future Energy Systems}, 2019.

\bibitem{Ember2026}
Ember, ``European wholesale electricity price data.'' Online, 2026.
\newblock Available at
  \url{https://ember-energy.org/data/european-wholesale-electricity-price-data/}
  [Accessed 21-02-2026].

\bibitem{NED2026}
{Nationaal Energie Dashboard}, ``Totale elektriciteitsproductie.'' Online,
  2026.
\newblock Available at \url{https://ned.nl}, [Accessed 21-02-2026].

\bibitem{Belastingdienst2026}
Belastingdienst, ``Energiebelasting.'' Online, 2026.
\newblock Available at \url{https://www.belastingdienst.nl} [Accessed:
  18-05-2026].

\bibitem{OpenEI2026}
{OpenEI}, ``Farm solar array.'' Online, 2026.
\newblock Available at
  \url{https://openei.org/wiki/PVDAQ/Sites/Farm\_Solar\_Array} [Accessed
  21-02-2026].

\bibitem{IEA2025}
IEA, ``Electricity 2025,'' tech. rep., {International Energy Agency}, 2025.

\bibitem{RangelovaJones2025}
K.~Rangelova and D.~Jones, ``How cheap is battery storage?,'' tech. rep.,
  Ember, 2025.

\bibitem{LazardPower2025}
{Lazard Power, Energy \& Infrastructure Group}, ``{LCOE+},'' tech. rep.,
  Lazard, 2025.

\bibitem{JasperZhouEtAl2026}
F.~B. Jasper, Y.~Zhou, E.~H., M.~J. Baumann, J.~Peters, D.~H. Neuhaus, and
  M.~Weil, ``Life cycle assessment of grid-scale battery storage: evaluating
  the environmental competitiveness of sodium-ion systems,'' {\em Energy
  Advances}, vol.~5, no.~4, pp.~550--570, 2026.

\bibitem{SutiknoArsadiandoEtAl2025}
T.~Sutikno, W.~Arsadiando, and H.~S. Purnama, ``Battery types and recent
  developments for energy storage in electric vehicles: technical criteria and
  battery management system,'' {\em Clean Energy}, vol.~9, no.~6, pp.~293--322,
  2025.

\bibitem{KnehrKubalEtAl2025}
K.~Knehr, J.~Kubal, M.~Effat, and S.~Ahmed, ``Estimated cost of {EV} batteries:
  2019-2025,'' tech. rep., Argonne National Laboratory, 2025.

\bibitem{ClementeSalazarEtAl2025}
M.~Clemente, P.~Maharjan, M.~Salazar, and T.~Hofman, ``Meta-analysis of life
  cycle assessments for li-ion batteries production emissions,'' {\em
  {International Journal of Life Cycle Assessment}}, vol.~30, pp.~2625--2641,
  2025.

\bibitem{IEA2024}
IEA, ``Global co2 emissions from transport by sub-sector in the net zero
  scenario, 2000--2030,'' 2024.
\newblock {Available online at:}
  \url{https://www.iea.org/data-and-statistics/charts/global-co2-emissions-from-transport-by-sub-sector-in-the-net-zero-scenario-2000-2030-2}.

\bibitem{TetteroVerdegaalEtAl2022}
W.~Tettero, B.~Verdegaal, A.~v. Zante, and H.~J. Idema, ``Kosten
  laadinfrastructuur logistiek laden op privaat terrein,'' tech. rep.,
  Nationale Agenda Laadinfrastructuur, 2022.

\bibitem{GarciaMarquesEtAl2014}
R.~Garcia, P.~Marques, and F.~Freire, ``Life-cycle assessment of electricity in
  {Portugal},'' {\em {Applied Energy}}, vol.~134, pp.~563--572, 2014.

\bibitem{TurconiSimonsenEtAl2014}
R.~Turconi, C.~G. Simonsen, I.~P. Byriel, and T.~Astrup, ``Life cycle
  assessment of the {Danish} electricity distribution network,'' {\em The
  International Journal of Life Cycle Assessment}, vol.~19, pp.~100--108, 2014.

\bibitem{ChatzipanagiJaeger-WaldauEtAl2024}
A.~Chatzipanagi, A.~{Jaeger-Waldau}, S.~Letout, A.~Mountraki, J.~G. Bermudez,
  A.~Georgakaki, E.~Ince, and A.~Schmitz, ``Clean energy technology
  observatory, photovoltaics in the european union – status report on
  technology development, trends, value chains and markets – 2024,'' tech.
  rep., European Commission. Joint Research Centre, 2024.

\bibitem{IRENA2025}
{IRENA}, ``Renewable power generation costs in 2024,'' tech. rep.,
  International Renewable Energy Agency, 2025.

\bibitem{KostMuellerEtAl2024}
C.~Kost, P.~M{ü}ller, J.~Sepúlveda~Schweiger, V.~Fluri, and J.~Thomsen,
  ``Levelized cost of electricity: Renewable energy technologies,'' tech. rep.,
  Fraunhofer Institute for Solar Energy Systems {ISE}, 2024.

\bibitem{Hernandez-LopezVega-De-LilleEtAl2024}
D.-A. Hernández-López, M.~I. Vega-De-Lille, J.~C. Sacramento-Rivero,
  C.~Ponce-Caballero, A.~El-Mekaoui, and F.~Navarro-Pineda, ``Life cycle
  assessment of photovoltaic panels including transportation and two
  end-of-life scenarios: Shaping a sustainable future for renewable energy,''
  {\em Renewable Energy Focus}, vol.~51, p.~100649, 2024.

\bibitem{LanzNollEtAl2022}
L.~Lanz, B.~Noll, T.~S. Schmidt, and B.~Steffen, ``Comparing the levelized cost
  of electric vehicle charging options in {Europe},'' vol.~13, no.~5277, 2022.

\bibitem{Dubois2024}
J.~Dubois, ``Life cycle assessment of electric vehicle charging infrastructure:
  systematic literature review and development of an life cycle inventory
  database,'' tech. rep., CIRAIG, 2024.

\end{thebibliography}

\ifextendedversion
\begin{appendices}
\section{Case Study Parameters}
This appendix lists the parameters used in the case study of Section~\ref{ch:Case_study}. Table~\ref{tab:Optimisation_parameters_fairness} reports the main technical parameters of the  hub components, the main statistics of the charging data~\cite{LeeLiEtAl2019} and the optimisation settings. Table~\ref{tab:Cost_parameters_fairness} reports the economic and emission parameters used to compute the amortisation costs in Section~\ref{ch: energy hub model}, together with their sources. For each component, the economic lifetime cost is $C^\mathrm{eco}_\mathrm{life} = \mathrm{CAPEX} + \mathrm{OPEX}\cdot l$, and the emission lifetime cost $C^\mathrm{GHG}_\mathrm{life}$ is its global warming potential obtained from life-cycle assessments.

\begin{table}[tbp]
\centering
\caption{University campus case study parameters}
\label{tab:Optimisation_parameters_fairness}
\begin{tabular}{l|l|r|l}
Parameter & Description & Value & Unit \\ \hline
\textbf{Components} &&\\
$E_{\mathrm{b}}^\mathrm{inst}$  & BESS capacity & 100 & kWh \\
$\overline{E}_{\mathrm{b}}$     & Max. BESS energy   & 90   & kWh  \\
$\underline{E}_{\mathrm{b}}$    & Min. BESS energy   & 10   & kWh  \\
$\overline{P}_{\mathrm{b}}$     & Max. BESS power            &  10  & kW   \\
$\underline{P}_{\mathrm{b}}$    & Min. BESS power            & $-10$  & kW   \\
$\overline{P}_{\mathrm{g}}$     & Max. grid power            &  150 & kW   \\
$\underline{P}_{\mathrm{g}}$    & Min. grid power            & $-150$ & kW   \\
$\overline{P}_{\mathrm{ch}}$ & Max. EV charging power     & 11   & kW   \\
$\underline{P}_{\mathrm{ch}}^{\mathrm{V2G}}$ & Min. EV charging power & $-11$  & kW    \\
$\underline{P}_{\mathrm{ch}}^{\mathrm{V1G}}$ & Min. EV charging power & 0  & kW  \\
\textbf{Datasets} &&\\
$P_{\mathrm{pv}}^\mathrm{inst}$ & Installed PV capacity & 21  & $\mathrm{kW_p}$  \\
& Avg. number of EVs & 29.1 & $\textrm{day}^{-1}$ \\
& Avg. EV connection time & 6.06 & h\\
& Avg. EV charging demand & 8.92 & kWh\\
\textbf{Optimisation} &&\\
$\Delta t$ & Time step resolution & 0.25 & h\\
$\alpha_\mathrm{r}$ & Regularisation weight & $1\cdot10^{-6}$ \\

\end{tabular}
\end{table}

\begin{table}[tbp]
\centering
\caption{University campus case study cost parameters}
\label{tab:Cost_parameters_fairness}
\begin{tabular}{l|r|l|l}
Parameter &  Value & Unit & Source \\ \hline

\textbf{BESS} &&&\\
$C_\mathrm{b, life}^{\mathrm{eco}}$ & 250 & $\text{€}/\mathrm{kWh}$ & \cite{RangelovaJones2025,LazardPower2025}\\
$C_\mathrm{b, life}^{\mathrm{GHG}}$ & 130 & kg $\mathrm{CO_2e}/\mathrm{kWh}$ & \cite{JasperZhouEtAl2026} \\
$l_\mathrm{b}$ & 10000 & cycles & \cite{RangelovaJones2025}\\
$a_\mathrm{b}^\mathrm{eco}$ & 1.56 & $\mathrm{ct}/\mathrm{kWh}$ & \eqref{eq:C_amort_bat} \\
$a_\mathrm{b}^\mathrm{GHG}$ & 8.1 & g $\mathrm{CO_2e}/\mathrm{kWh}$ & \eqref{eq:C_amort_bat} \\

\textbf{EV batteries} &&&\\
$C_\mathrm{evb, life}^{\mathrm{eco}}$ & 120 & $\text{€}/\mathrm{kWh}$ & \cite{SutiknoArsadiandoEtAl2025,KnehrKubalEtAl2025} \\
$C_\mathrm{evb, life}^{\mathrm{GHG}}$ & 80 & kg $\mathrm{CO_2e}/\mathrm{kWh}$ & \cite{ClementeSalazarEtAl2025,IEA2024} \\
$l_\mathrm{evb}$ & 5000 & cycles & \cite{SutiknoArsadiandoEtAl2025} \\
$a_\mathrm{evb}^\mathrm{eco}$ & 1.2 & $\mathrm{ct}/\mathrm{kWh}$ & \eqref{eq:C_amort_bat} \\
$a_\mathrm{evb}^\mathrm{GHG}$ & 8 & g $\mathrm{CO_2e}/\mathrm{kWh}$ & \eqref{eq:C_amort_bat} \\

\textbf{Grid} &&&\\
$\mathrm{CAPEX}_\mathrm{g}$ & 70 & $\text{€}/\mathrm{kW}$ & \cite{TetteroVerdegaalEtAl2022} \\
$\mathrm{OPEX}_\mathrm{g}$  & 27 & $\text{€}/\mathrm{kW}/\textrm{year}$ & \cite{TetteroVerdegaalEtAl2022} \\
$l_\mathrm{g}$ & 40 & years & 
\cite{TetteroVerdegaalEtAl2022,GarciaMarquesEtAl2014,TurconiSimonsenEtAl2014} \\
$a_\mathrm{g}^\mathrm{eco}$ & 4.6 & $\mathrm{ct}/\mathrm{kWh}$ & \eqref{eq:grid_amort_price} \\
$a_\mathrm{g}^\mathrm{GHG}$ & 1.84 & g $\mathrm{CO_2e}/\mathrm{kWh}$ & \cite{GarciaMarquesEtAl2014,TurconiSimonsenEtAl2014} \\
Energy tax & 5.548 & $\mathrm{ct}/\mathrm{kWh}$ & \cite{Belastingdienst2026} \\

\textbf{PV} &&&\\
$\mathrm{CAPEX}_\mathrm{pv}$ & 800 & $\text{€}/\mathrm{kW_p}$ & 
\cite{ChatzipanagiJaeger-WaldauEtAl2024, IRENA2025, KostMuellerEtAl2024} \\
$\mathrm{OPEX}_\mathrm{pv}$  & 9 & $\text{€}/\mathrm{kW_p}/\textrm{year}$ & \cite{ChatzipanagiJaeger-WaldauEtAl2024, IRENA2025, KostMuellerEtAl2024} \\
$C_\mathrm{pv, life}^{\mathrm{GHG}}$ & 593 & kg $\mathrm{CO_2e}/\mathrm{kW_p}$ & \cite{Hernandez-LopezVega-De-LilleEtAl2024} \\
$l_\mathrm{pv}$ & 25 & years & 
\cite{IRENA2025,Hernandez-LopezVega-De-LilleEtAl2024} \\
$a_\mathrm{pv}^\mathrm{eco}$& 41 & $\text{€}/\mathrm{kW_p}/\textrm{year}$ & \eqref{eq:C_amort_PV} \\
$a_\mathrm{pv}^\mathrm{GHG}$ & 23.7 & kg $\mathrm{CO_2e}/\mathrm{kW_p}/\textrm{year}$ & \eqref{eq:C_amort_PV} \\

\textbf{Chargers} &&&\\
$\mathrm{CAPEX}_\mathrm{ch}$ & 3000 & $\text{€}/\mathrm{charger}$ & \cite{TetteroVerdegaalEtAl2022,LanzNollEtAl2022} \\
$\mathrm{OPEX}_\mathrm{ch}$  & 100 & $\text{€}/\mathrm{charger}/\textrm{year}$ & \cite{TetteroVerdegaalEtAl2022,LanzNollEtAl2022} \\
$C_\mathrm{ch, life}^{\mathrm{GHG}}$ & 622 & kg $\mathrm{CO_2e}/\mathrm{charger}$ & \cite{Dubois2024} \\
$l_\mathrm{ch}$ & 15 & years & 
\cite{LanzNollEtAl2022} \\
$N_\mathrm{ch}$ & 40 & chargers & \\
$a_\mathrm{ch}^\mathrm{eco}$ & 12.7 & $\mathrm{ct}/\mathrm{kWh}$ & \eqref{eq:chargers_amort_price} \\
$a_\mathrm{ch}^\mathrm{GHG}$ & 17.5 & g $\mathrm{CO_2e}/\mathrm{kWh}$ & \eqref{eq:chargers_amort_price} \\

\end{tabular}
\end{table}

\end{appendices}
\fi

\end{document}